\documentclass[a4paper]{jpconf}
\usepackage{graphicx}
\begin{document}
\title{Neutrino emission from magnetized micro-quasar jets}

\author{Theodoros Smponias$^1$ and Odysseas Kosmas$^2$}

\address{$^1$Division of Theoretical Physics, University of Ioannina, GR-45110 Ioannina, Greece}
\address{$^2$Modelling and Simulation Centre, MACE, University of Manchester, Sackville Street, Manchester, UK}

\ead{t.smponias@hushmail.com, odysseas.kosmas@manchester.ac.uk}

\begin{abstract}
The hadronic jets in a micro-quasar stellar system are modelled with 
the relativistic hydrocode PLUTO. We focus on neutrino emission from 
such jets produced by fast proton (non-thermal) collisions on thermal 
ones within the hadronic jet. We adopt a semi-analytical approximation 
for the description of the secondary particles produced from p-p collisions 
and develop appropriate algorithms using the aforementioned 
injected protons as input. 
As a concrete example we consider the SS-433 X-ray binary system for which 
several observations have been performed the last decades. In contrast to 
the pre-set distribution of the fast protons along the jet employed in our 
previous works, in the present paper we simulated it by using a power-law 
fast proton distribution along the PLUTO hydro-code. This distribution 
gradually sweeps aside the surrounding winds, during the jet advance through the computational grid. As a first step, in the present work the neutrino energy spectrum is extracted from the model jet, facilitating a range of potential dynamical simulations in currently interesting microquasar jet systems.
\end{abstract}

\section{Introduction}

In binary stars commonly known as micro-quasars (MQs), two oppositely emitted 
jets of matter and radiation are produced. These systems are similar to 
Active Galactic Nuclei (AGN or quasars) and consist of  a main sequence star 
(the giant companion or donor star), in coupled orbit with a compact 
astrophysical object (a neutron star or a black hole) \cite{Mirabel_1999}. 
A characteristic 
mass accretion disk develops close to the compact object from mass absorption 
through the inner Lagrangian Point (Roche Lobe Overflow) due to angular momentum 
conservation. The jets of a MQ appear quite collimated (due to the presence of
a rather strong magnetic field) forming a multi-wavelength 
and also particle emitter \cite{Romero_2003,Reynoso_2008,Reynoso09}. 

Stellar MQs are currently important astrophysical systems with growing interest in 
their investigations within astrophysics, particle physics and cosmology. In the 
case of black hole micro-quasars (when the compact object is a black hole) the
stellar system provides excellent testing grounds for black hole theories. Therefore, 
an improved understanding of the dynamical astrophysical conditions within the jets 
in MQs is of significant importance \cite{Ferrari_1998,Fabrika_2004,Smponias_2015}.

In hadronic micro-quasar jets, the proton-proton interactions with the subsequent decays 
of the secondary particles, mostly $\pi^{\pm}$ mesons, produce high energy neutrinos. 
These collisions result also in the production of high-energy gamma rays, through the 
neutral pion ($\pi^0$) decay, as discussed in previous works \cite{Smponias_2015, Actis,Fermi,tsk-advanc-2015,Kelner_2006}. Recent simulations of high energy p-p 
interactions in terrestrial laboratories provide quite accurate energy distributions of secondary products in the high energy range (above 100 GeV) and 
determine parametric expressions of energy spectra for secondary particles 
like $\pi^{0}$ and $\pi^{\pm}$ mesons, neutrinos but also for gamma rays and electrons produced in inelastic p-p collisions \cite{Kelner_2006,Lipari07}. Such distributions may also be implemented when studying the hadronic MQs as neutrino and gamma ray sources \cite{Smponias_2015}.

Among the hadronic models proposed for the energy emission from micro-quasars (MQs), 
two are the most important: (i) In the first, relativistic protons in the jet interact 
with target protons from the stellar wind of the companion star. (ii) In the second,
neutrinos and gamma-rays are produced from p-p interactions between relativistic 
(non-thermal) and cold (thermal) protons within the jets themselves
\cite{Romero_2003,Reynoso_2008,Reynoso09,Kelner_2006,Lipari07}. In the latter 
case, relativistic (fast) protons within the jet are subject to different mechanisms 
that can make them lose energy. It is interesting to know the energy range where p-p
collisions are the main (dominant) cooling process that produces the corresponding
neutrinos (or gamma-rays). On the other hand, the cold (slow) protons serve as targets 
for the relativistic protons \cite{PS_2001,TR_2011}.

From a phenomenological point of view, micro-quasar neutrino and gamma ray sources need 
to be modelled fully relativistically \cite{Mirabel_1999,Smponias_2015}. A suitable treatment 
is offered by the relativistic hydrocodes developed recently, such as the relativistic 
magnetohydrodynamical (RMHD) PLUTO hydro-code \cite{Mignone_2007} employed in Ref.
\cite{Smponias_2015,Smponias_2011,Smponias_2014} in order to simulate the hadronic jets 
of the SS-433 MQ, an X-ray binary star \cite{Ferrari_1998,Fabrika_2004,Margon_1984}.

The present paper, is an extension of our work of Ref. \cite{Smponias_2015} where we 
modeled simulated neutrino emission from Galactic astrophysical hadronic jets originating 
from the vicinity of compact objects in binary stellar systems. Our dynamical simulations
come out of the RMHD PLUTO code in conjunction with the in house developed (in C,
Mathematica and IDL) codes. We now produce further results 
that aim to be directly comparable to the sensitivities of modern high energy neutrino detectors, 
e.g.~the IceCube \cite{IceCube} and KM3NeT \cite{KM3Net},
thus clarifying the potential for observing neutrino emissions from micro-quasars. 

\section{Brief description of the main background and formalism}

In this work, we adopt the model explaining the neutrino and gamma-ray production 
through the p-p interactions between relativistic and cold protons occuring within the 
MQ jets themselves \cite{Romero_2003,Reynoso_2008,Reynoso09,Kelner_2006,Lipari07}. 
Relativistic protons in the jet are subject to various mechanisms that can lead to
energy release. As is well known, in the case of hadronic MQ jets, a small portion 
(about $1\%$) of the protons (bulk flow protons) may be accelerated through first 
order Fermi acceleration procedures that take place essentially at shock fronts 
inside the jet. In general, accelerated particles within the jet may gain energy up 
to the $TeV$ scale. 

For the particle (proton) acceleration rate at shocks (first order Fermi mechanism), 
we have
\begin{equation}
t^{-1}_{acc} \simeq \eta \frac{c e B}{E_{p}},
\end{equation}
where $B$ denotes the magnetic field and $E_p$ the proton energy ($e$ and $c$ are the 
usual parameters, i.e.~the proton charge and the speed of light, respectively). The 
acceleration efficiency parameter $\eta$ in our present calculations is set equal to 
$\eta=0.1$ (efficient accelerator case, mildly relativistic shocks near the jet base) 
\cite{Begelman-1980}.

From the scattering of fast protons off slow protons, high energy pions and kaons are 
produced which may further decay to very high energy gamma rays and neutrinos. The 
reaction schemes are described by equations of the form
\begin{equation}
pp \rightarrow pp \pi^{0}+F_{0} \, ,
\end{equation}
for the neutral-pion ($\pi^{0}$) production channel, and
\begin{equation}
pp \rightarrow pn \pi^{+}+F_{1} \, , \qquad pp \rightarrow pn \pi^{-}+F_{2} \, ,
\end{equation}
for the charged-pion ($\pi^{\pm}$) production channels, where $F_{i}, \, \, i={0,1,2}$, 
comprises $\pi^{0}$ and $\pi^{+}\pi^{-}$ pairs, respectively. 

Subsequently, the neutral pions $\pi^0$ and other neutral mesons decay quickly producing 
high energy gamma-rays. The charged pions $\pi^+$ ($\pi^-$), needed for the purposes of 
our present work (also the charged-kaons), decay and lead to muons and furthermore to 
the production of various flavors of neutrinos as discussed below.

\subsection{Secondary charged particle decays}

From inelastic p-p scatterings among non-thermal protons and thermal ones within the 
hadronic jet, neutrinos are mainly produced through charged pion decays (known as prompt 
neutrinos). The muons included in the by-products can afterwards decay again into an 
electron (or a positron) and the associated two light neutrino flavors (delayed neutrino 
beam) according to the reactions described below. 

\subsubsection{Prompt decay channels (prompt neutrinos):}

The $\pi^{+}$ ($\pi^{-}$) mesons (with a mass of $m_\pi = 139.6 MeV/c^{2}$ and a half-life 
of $2.6×10^{-8}$s) decay due to the weak interaction, the primary decay mode of which
(with a probability of $0.999877$) is a reaction leading to an anti-muon (muon) and a 
muonic neutrino (muonic anti-neutrino) as
\begin{equation}
\pi^{+} \rightarrow	\mu^{+} + \nu_{\mu} \, ,  \qquad 
\pi^{-} \rightarrow \mu^{-} + \widetilde{\nu}_{\mu} \, .
\label{muon-produc}
\end{equation}
A less important decay mode of a $\pi^{+}$ ($\pi^{-}$), with probability of occurrence 
just $0.000123$, is its decay into a positron (electron) and an electron neutrino 
(electron anti-neutrino) as
\begin{equation}
\pi^{+} \rightarrow	e^{+} + \nu_{e} \, ,  \qquad 
\pi^{-} \rightarrow e^{-} + \widetilde{\nu}_{e} \, .
\end{equation}
In this work we neglect the neutrino production through the latter channels.

\subsubsection{Delayed decay channel (delayed neutrinos):}

The other important source of neutrinos in hadronic jets is the decay mode of the 
produced muons (muon leptonic decays) in the reactions (\ref{muon-produc}), which 
produces also two neutrinos described by the processes
\begin{equation}
\mu^{+} \rightarrow  e^{+}  + \nu_{e}\, + \widetilde{\nu}_{\mu} \, ,  \qquad 
\mu^{-} \rightarrow  e^{-} + \widetilde{\nu}_{e} +  \nu_{\mu} \, .
\end{equation}

In general, the analytical formulae 
suggested from laboratory p-p collisions resemble the simulated 
distributions extracted in Ref. \cite{Kelner_2006} within a few percent over a large range of the fraction of the energy of 
the incident proton ($E_p$) transferred to the secondary particles, i.e.~the ratio 
$x = E_i/E_p$, with $E_i$ being the energy of the secondary particle (e.g.~pion).

From an experimental point of view, for astrophysical gamma rays and neutrinos extremely 
sensitive detection systems have been developed \cite{Actis,Fermi,KM3Net}. These detectors sparked 
a renewed interest for studying stellar objects as neutrino and gamma ray sources, e.g.~the 
SS-433 system is widely known from the early 80's as the only MQ with a verified hadronic jet 
content. We mention, for example, that observations of iron lines in the spectrum of the 
SS-433 MQ provided useful information regarding the hadronic content of its jets 
\cite{Begelman-1980}.  

From a theory and phenomenology point of view, the gamma ray and neutrino production from 
a hadronic MQ that are of interest in the present work, is based on reliably determining 
the distribution of the fast protons and the realistic injection functions of the produced 
secondary particles (pions, kaons, muons, etc.).
 
In previous works \cite{Smponias_2015,Smponias_2011,Smponias_2014}, the hadronic jet was 
modelled using the PLUTO code. The results of PLUTO were then processed in order to calculate 
the emissivity of various  secondary particles (pions, kaons) and the produced muons, gamma-rays, 
etc., on the basis of the spatial and time variation of physical parameters like the magnetic field 
that collimated the jet, the mass number density for every grid cell of the PLUTO code and others.

Before proceeding to the presentation and discussion of the results we should mention that,
the discrimination of prompt and delayed neutrinos from MQ jets is not possible, therefore the 
results obtained in the present work refer to physical quantities pertaining to prompt neutrinos, 
nonetheless as they are much faster to simulate computationally.

\section{Results and discussion}

The main results of this work refer to the mean number density of the non-thermal protons
(obtained with the algorithms mentioned before and the PLUTO hydrocode), the pion injection 
function and the pion energy distribution describing the pion governing Eq.~(\ref{muon-produc}).
The evaluation of the emissivity of the prompt neutrinos relies on these calculations.   

\subsection{Non-thermal proton density}

We begin our calculations by considering the production of non-thermal protons in the jet. 
The non-thermal proton population emerges from the bulk jet flow that comprises mainly thermal 
protons, moving mildly relativistically. Some of the slow protons are locally accelerated, at 
shock fronts appearing within the jet flow (first-order Fermi acceleration process), to 
ultra-relativistic velocities. While in our previous studies we adopted a fast (non-thermal) 
proton jet density N$_{p}$, equal to a tiny fraction (10$^{-6}$) of the corresponding thermal proton 
density, in the present work, we assume a power-law distribution of the form $N_p = N_0 E^{-\alpha}$, 
with $\alpha \approx 2$ \cite{Reynoso_2008}.  In addition, we considered a spatial density 
distribution n$_{z}$, coming out of explicit calculations with the PLUTO hydro-code as discussed below.

\begin{figure}[!htb]
\begin{center}
\vspace*{-3.3 cm}
\includegraphics[width=0.65\textwidth]{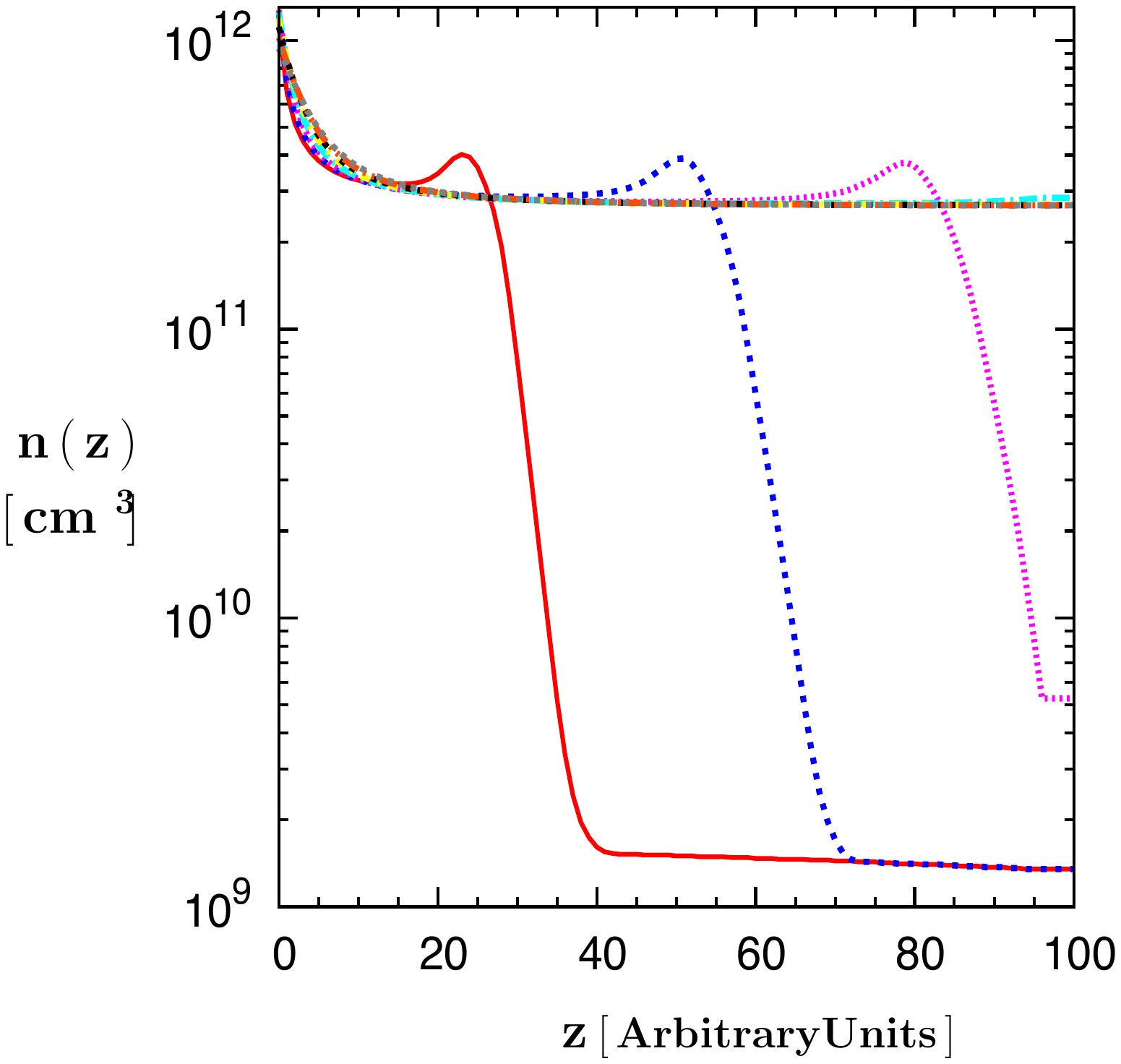}
\vspace*{-3.7 cm}
\caption{Density profile evolution for slow protons along the jet. Each snapshot represents 
100 PLUTO time units in the simulation, or $33$ seconds in model time. The beginning rests 
at the start of the simulation. The first three or four snapshots indicate a dynamic jet 
evolution, while the rest approach a steady-state behaviour.}
\end{center}
\label{fig-1}
\end{figure}

For an RMHD simulation of a rather laterally restricted magnetized jet we, first, calculated 
the mean matter density along the jet axis (as a function of z), i.e.~the slow proton density $n(z)$, by evaluating the PLUTO density over a slice cut perpendicular to the jet axis. In order to cover the temporal evolution of the jet as the simulation evolves, these mean density values have been obtained for a number of $8$ snapshots which are plotted in Fig.~1. From this Figure we can see how the mean density profile evolves along the jet. Its peak is moving outwards while the overall maximum gradually decreases. The jet remains confined, mainly due to the presence of a toroidal magnetic field component (B$_{tor}$). The surrounding wind helps shape the jet as well, especially at the early stages of the simulation, before the wind begins to be swept by the jet.  

As the jet advances through the computational grid, it gradually sweeps aside the surrounding 
winds resulting to a near-steady-state with a rather flat density profile. The magnetic jet 
confinement prevents the jet density from falling too much along the jet. It is worth mentioning
that, for the characteristic time scales of the energy loss mechanisms, we largely follow 
Refs.~\cite{Kelner_2006} and \cite{Reynoso09}, incorporating mainly synchrotron and adiabatic 
energy loss mechanisms.

\subsection{Pion injection function and pion energy distribution}


For every p-p interaction (one 'fast', non-thermal proton scattered off a 'slow', thermal 
one) we obtain a probability density of a resulting pion at every position along the possible 
spectrum of resulting pions, i.e.~we get a spectrum of possible energies for the resulting pion. 
That spectrum, per p-p collision, is represented by $F_{\pi}$ and is dependent on the incoming 
fast proton energy (slow proton Energy is negligible by comparison) and the ratio of a given 
position at the pion spectrum to the incoming proton energy. 
In Ref.~\cite{Kelner_2006} the function $F_{\pi}$  is given by the expression

\begin{eqnarray}
F_{\pi}^{(pp)}\left( x,\frac{E}{x} \right) &=& 4 \alpha B_{\pi} x^{\alpha-1} 
\left( \frac{1-x^{\alpha}}{1+r x^{\alpha}(1-x^{\alpha})} \right)^{4}  \\ \nonumber 
&&\left( \frac{1}{1-x^{\alpha}} + \frac{r(1-2x^{\alpha})}{1+rx^{\alpha}(1-x^{\alpha})} 
\right) \left( 1- \frac{m_{\pi} c^{2}}{x E_{p}}  \right)^{\frac{1}{2}} 
\end{eqnarray}
which represents the pion spectrum per proton-proton interaction. $x=E/E_{p}$, 
$B_{\pi}=a'+0.25$, $a'=3.67+0.83L+0.075L^{2}$, $r=2.6/ \sqrt{a'}, \alpha=0.98/\sqrt{a'}$,
and $L$ is the jet's luminocity (see \cite{Reynoso09,Kelner_2006}).
In Fig.~2 (left panel) the product $xF_{\pi}$ is plotted as a function of the ratio 
$x$, for three different incoming fast proton energies (E$_{p}$=10$^{3}$GeV, E$_{p}$=10$^{4}$GeV, 
E$_{p}$=10$^{5}$GeV), which cover the energy range of interest.

\begin{figure}[!htb]
\vspace*{-1.0 cm}
\begin{center}
\includegraphics[width=0.565\textwidth]{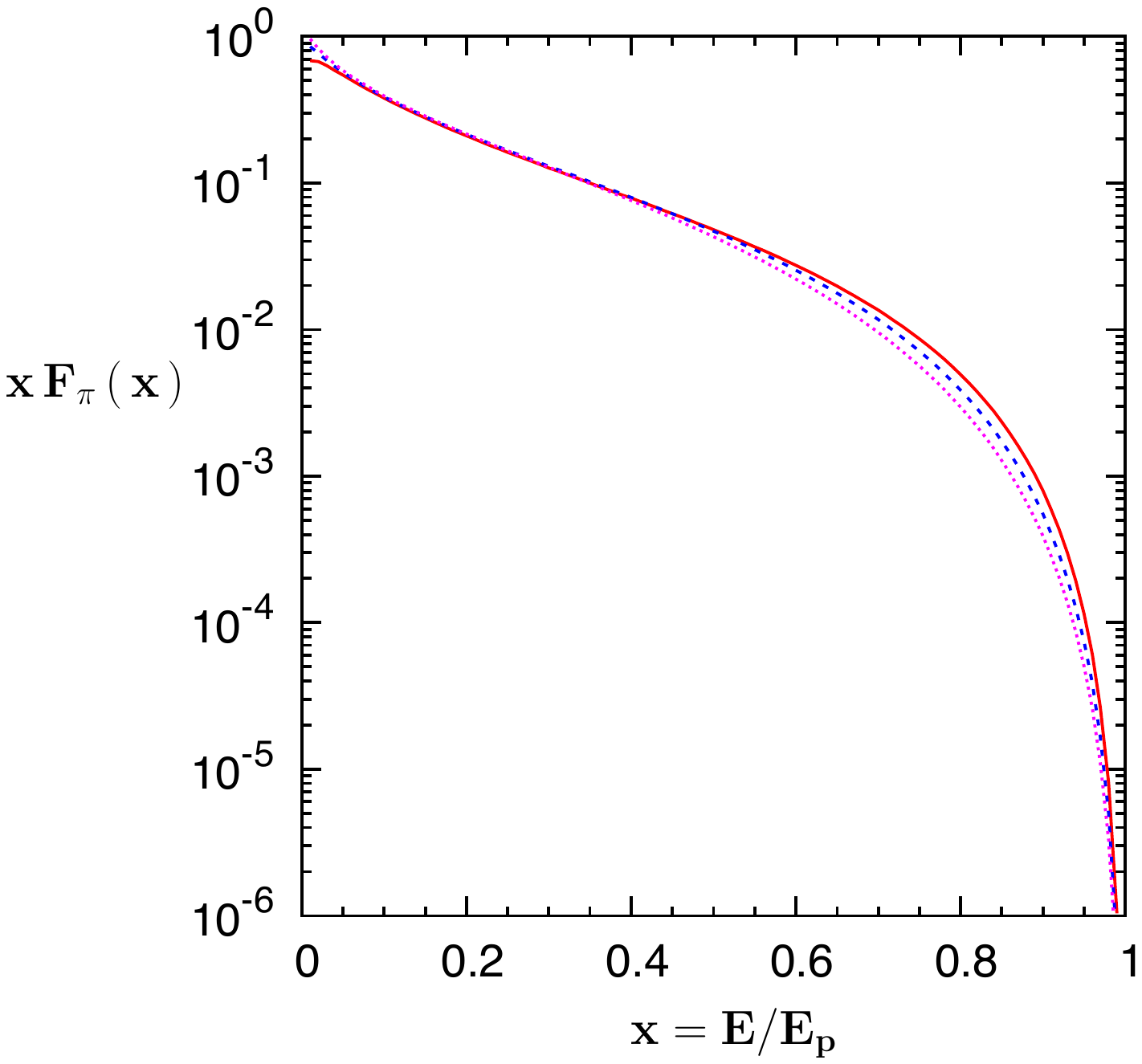}
\hspace*{-2.5 cm}
\includegraphics[width=0.565\textwidth]{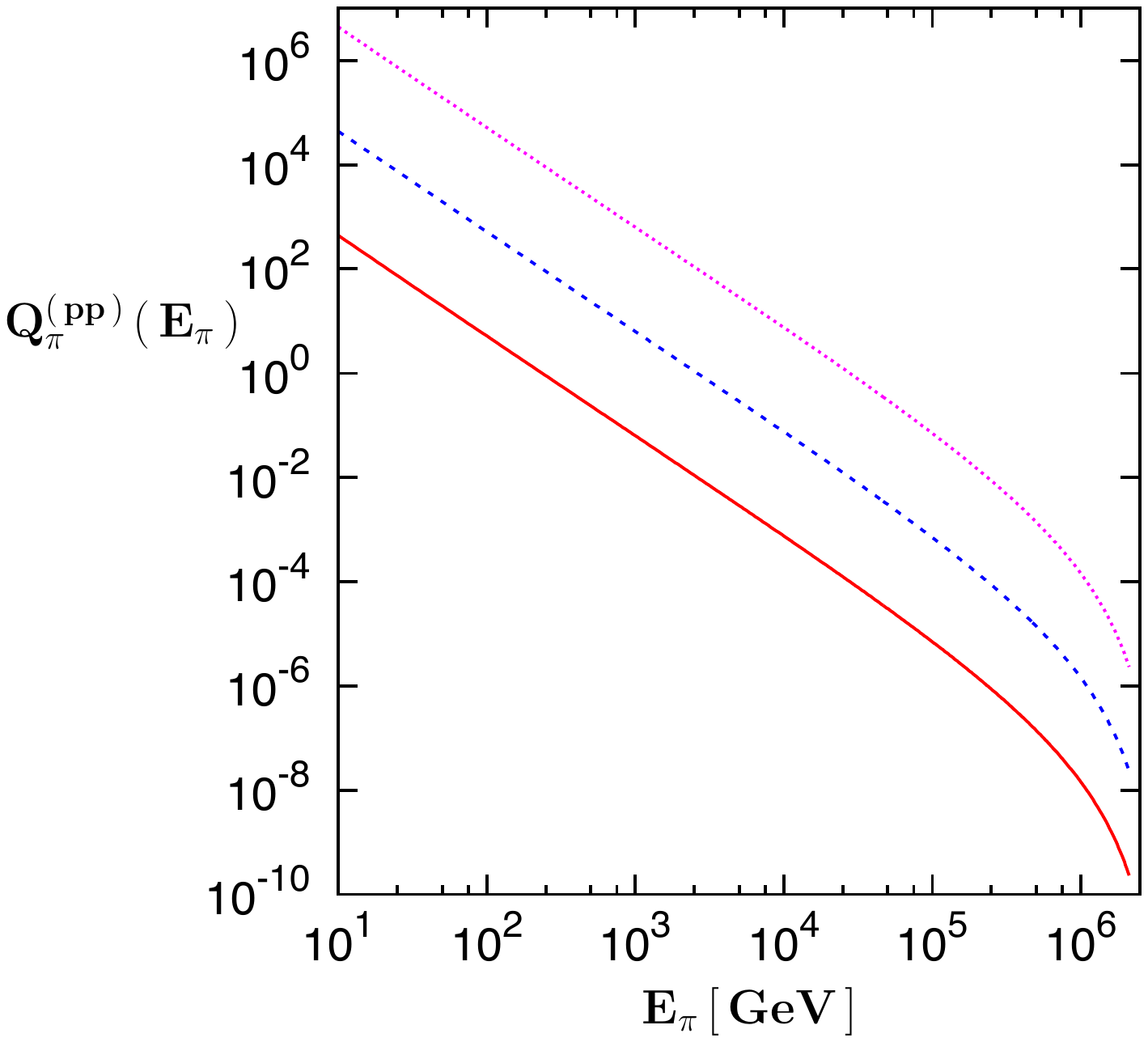}
\vspace*{-2.9 cm}
\caption{ Left: Pion energy spectra, per p-p collision, obtained for three different incoming fast proton energies (E$_{p}$=10$^{3}$ GeV, E$_{p}$=10$^{4}$ GeV, E$_{p}$=10$^{5}$ GeV), as a function of the ratio x=E$_{\pi}$/E$_{p}$. We consider charged pions ($\pi^{\pm}$) as needed for our purposes in this work. E$_{\pi}$ denotes the secondary particle (pion) energy. Right: Variation of the pion injection function, $Q_{\pi}^{(pp)}(E)$ through the pion energy spectrum, for three different jet densities n(z), located at different points along the model jet axis.}
\end{center}
\label{fig-2}
\end{figure}

With the aid of this function, we calculate the pion injection function, $Q_{\pi}^{(pp)}$
through the relation
\begin{eqnarray}
Q_{\pi}^{(pp)}(E,z) = n(z) c \int \limits_{k}^{1} \frac{dx}{x} N_{p}
\left(  \frac{E}{x},z \right) F_{\pi}^{(pp)} \left( x,\frac{E}{x} \right) 
\sigma^{(inel)}_{pp} \left( \frac{E}{x} \right)  \, ,
\label{Qpp}
\end{eqnarray}
where
$ k = E/ E_{p}^{(max)}$. N$_{p}$ stands for the fast proton density, $x$ is the ratio of 
the pion energy to proton energy, and $\sigma^{inel}_{pp}$ is the proton-proton inelastic 
collision cross section. 

The pion injection function, $Q_{\pi}^{(pp)}$ depends on the thermal proton density, 
$n(z)$. In Fig.~2 (right panel), we plot the $Q_{\pi}^{(pp)}$ versus the pion 
energy $E_\pi$ for three different jet densities (n=10$^{9}$, n=10$^{10}$, n=10$^{11}$). 
We notice the approximate square dependence of the scale of Q$_{\pi}^{(pp)}$ on the jet 
density, which is because N$_{p}$ also depends on $n(z)$.

As a physical interpretation, let us consider a large number of p-p collisions. So, we 
add up, at every pion spectrum energy, the contributions to the probability that a pion will result at that 
energy. Depending on the incoming proton energy for each collision, there may be a smaller or 
a larger contribution to any given pion energy, as long as it is smaller than the proton's 
energy in the first place (pion energy cannot exceed proton energy). So we integrate over 
many p-p collisions to find the pion spectrum of a collection of p-p collisions, i.e.~the 
pion injection function Q$_{\pi}^{(pp)}$ .

In order to obtain the pion distribution entering neutrino emissivity, we solve the 
following transport equation
\begin{equation}
\frac{\partial N_{\pi}}{\partial E} + \frac{N_{\pi}}{t_{loss}} = Q_{\pi}^{(pp)}(E,z)
\end{equation}
where N$_{\pi}(E,z)$ denotes the pion energy distribution. The numerical integration
of the transport equation, for a cell of the hydrocode, i.e.~a localized position in
space, is given by the following expression
\begin{eqnarray}
N_{\pi}(E) = \frac{1}{|b_{\pi}(E)|} \int \limits_{E}^{E^{(max)}} dE' Q_{\pi}^{(pp)}(E') 
\exp {[-\tau_{\pi}(E,E')]} \, ,
\end{eqnarray}
where 
\begin{eqnarray}
\tau_{\pi}(E',E)=\int \limits_{E'}^{E} \frac{dE'' t_{\pi}^{-1}(E)}{|b_{\pi}(E'')|} \, .
\end{eqnarray}
We note here that, the physical conditions within a cell are taken to be constant and 
also that the macroscopic physical parameters (density, pressure, etc) within each cell 
are taken to be constant. Under these assumptions, the transport equation is only 
dependent on energy, which considerably simplifies its calculation. We also take the 
characteristic scale (mean free path) of the radiative interactions to be smaller than 
the cell size, leading to the containment of particle interactions within a given 
hydrocode cell. Furthermore, the time scale for the radiative interactions is so much 
smaller than the hydrocode's timestep, that the radiative interactions belong to a 
single timestep each time.    

The behaviour of the pion distribution $N_{\pi}(E_p)$, in the energy range of our interest, 
is illustrated in Fig.~3. This curve refers to a typical computational cell of 
the PLUTO hydrocode. It could be easily extended to a number of hydrocode cells covering 
a span of the computational grid, therefore opening the way towards obtaining the neutrino 
emissivity from the whole grid. 

In such a treatment, we consider a large number of interacting particles per computational 
cell, therefore the probability density in the transport equation can be approximated by 
the number density of the particles, rendering inactive the stochastic portion of the general 
transport equation. Moreover, only the deterministic portion of the transport equation is 
employed, which simplifies it to a deterministic partial differential equation (for 
further details on the meaning of various symbols and functions used in this section, the 
reader is referred to \cite{Reynoso09,Kelner_2006}).  

\begin{figure}[!htb]
\begin{center}
\includegraphics[width=0.77\textwidth]{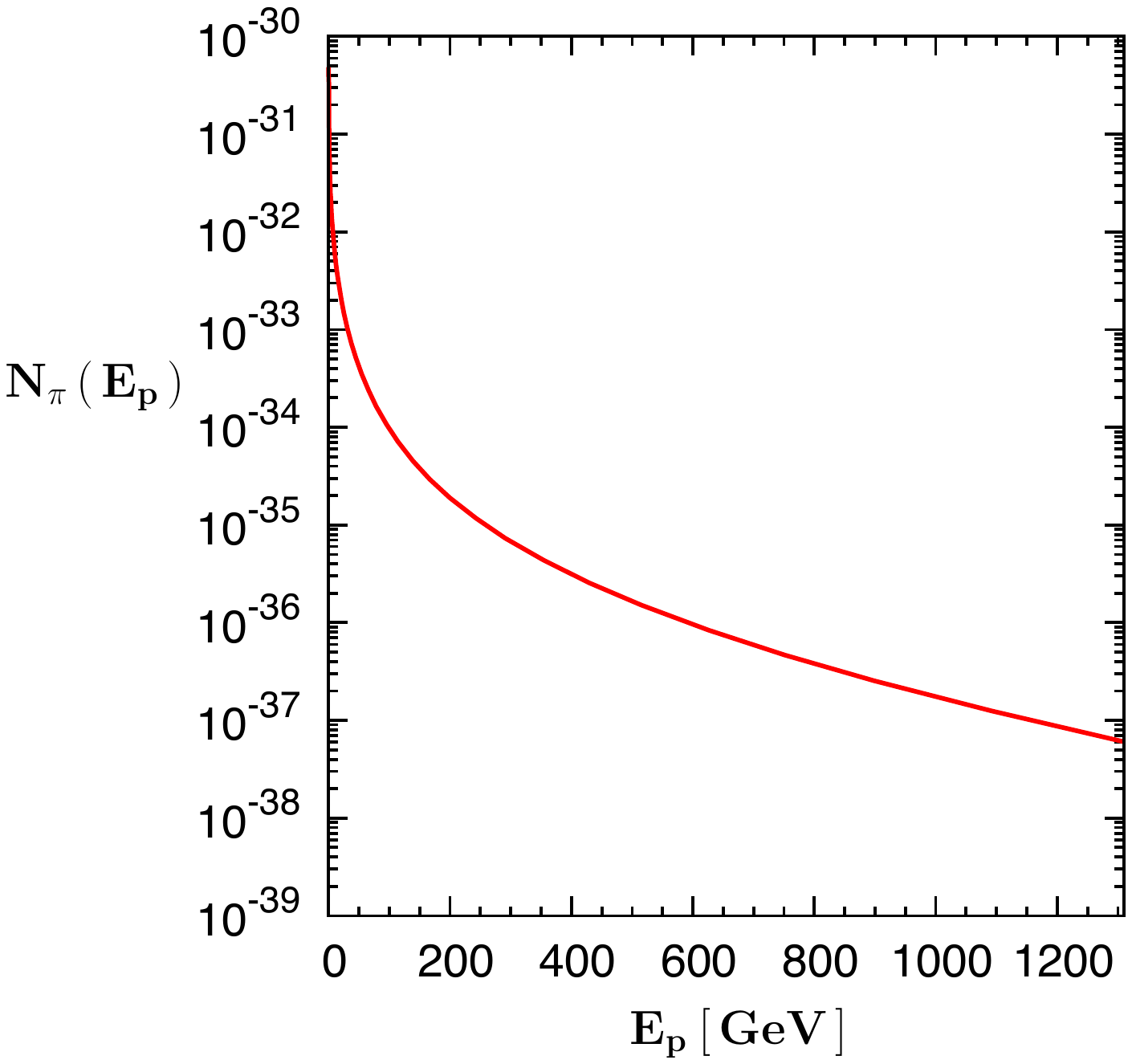}
\vspace*{-3.5 cm}
\caption{Pion energy distribution $N_{\pi}(E_p)$ obtained through
the solution of the transport equation, for a typical cell of the 
hydrocode. This cell is a localized position in the space of the jet.
\label{fig-3}}
\end{center}
\end{figure}

\subsection{Neutrino emissivity}

As mentioned before, in this work we consider neutrinos emanating from direct pion 
decays (prompt neutrinos, see reaction \ref{muon-produc}). In the semi-analytical 
approach implemented in this work, the emissivity of prompt neutrinos, is obtained 
with the aid of $N_{\pi}(E_p)$ from the expression \cite{Reynoso09,Lipari07}
\begin{eqnarray}
Q_{\pi \rightarrow \nu}(E) = \int \limits_{E}^{E_{max}} dE_{\pi} t^{-1}_{\pi,dec} 
(E_{\pi}) N_{\pi}(E_{\pi}) \frac{\Theta (1-r_{\pi}-x)} {E_{\pi}(1-r_{\pi})}  \, ,
\label{Neut-Emiss}
\end{eqnarray}
where $x=E/E_{\pi}$ and $t_{\pi,dec}$ is the  pion decay time-scale. $\Theta$($\chi$) 
is the well-known theta function (for further parameter details see Ref.~\cite{Smponias_2015}). 
The neutrino emission calculation could be performed mainly following the analysis 
of Refs. \cite{Reynoso_2008,Reynoso09,Kelner_2006,Lipari07}. 

For the readers' convenience, we should mention the following. The non-thermal proton
distribution suffers synchrotron and adiabatic losses, affecting the balance in the 
transport between protons and pions. 
The neutrino emissivity can then be integrated over $3$D cells volume (voxels volume) and divided by the surface of a sphere with radius the distance to Earth. The result is a synthetic 'neutrino emission observation' of the binary system. By repeating the process for many energies, we can then obtain a synthetic spectral emission distribution, for direct comparison with observations. 

\begin{figure}[!htb]
\begin{center}
\includegraphics[width=0.90\textwidth]{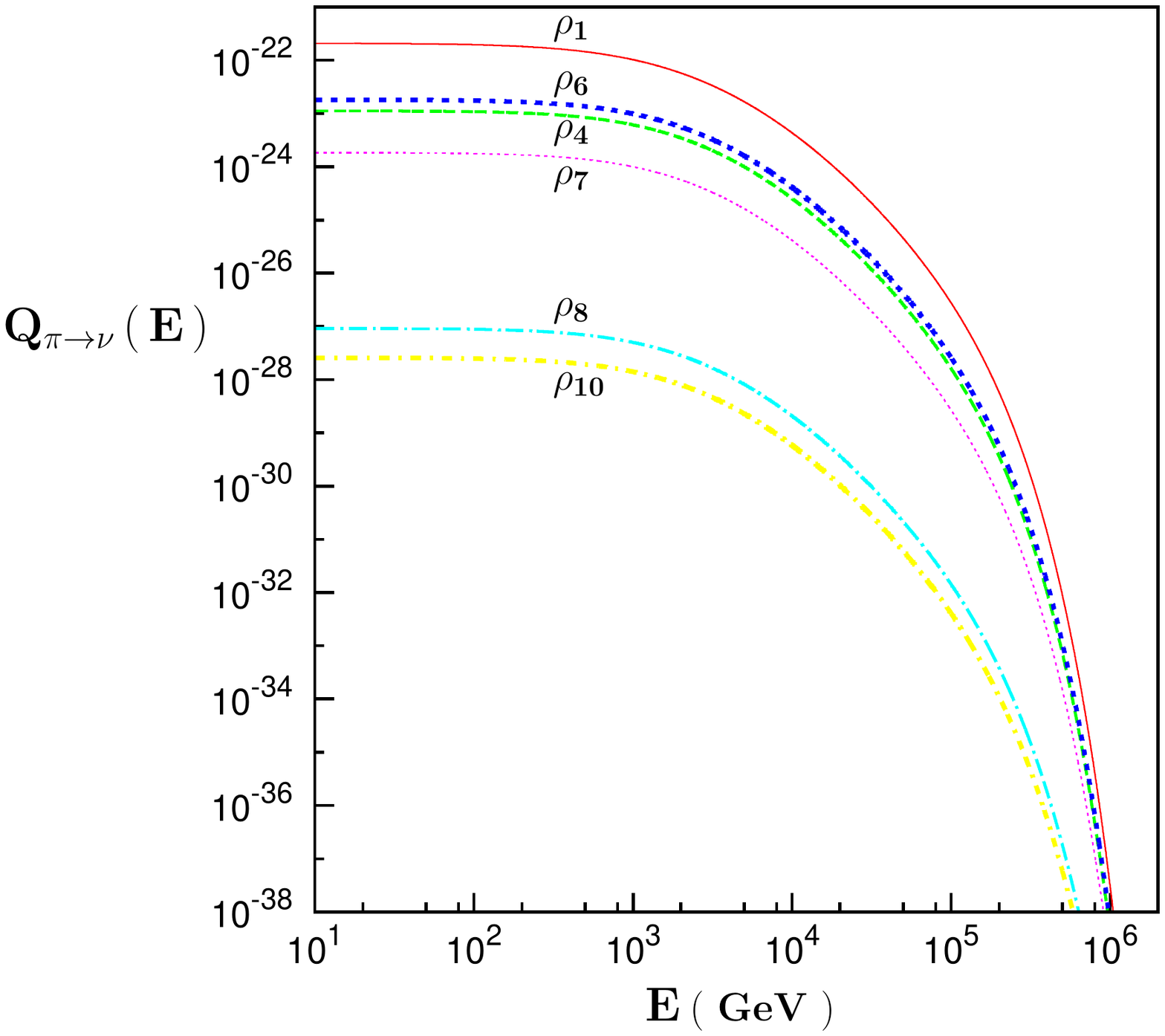}
\caption{ Neutrino emissivity $Q_{\pi \rightarrow \nu}(E)$ obtained for various
values of number density $\rho_{j}$ (see Table 1) by integrating numerically Eq. (\ref{Neut-Emiss}). For the sake of comparison with observations and other predictions, the value of $Q_{\pi \rightarrow \nu}(E)$ should be multiplied by the normalization factor F$_{norm}$=10$^{54}$ erg/GeV (see the text). 
\label{fig-4}}
\end{center}
\end{figure}
 
As an illustration of the behaviour of neutrino emissivity $Q_{\pi \rightarrow \nu}(E)$ versus the neutrino energy, in Fig. 4 the neutrino spectra from a series of computational slices, cut perpendicular to the jet axis (at equal intervals along the model jet), are shown. The density of each slice is spatially averaged over the slice surface and that average density (see Table 1) is then employed in the neutrino emission calculation. The averaging is performed in IDL and the emission calculation in Mathematica. The results presented in Fig. 4 are un-normalized, but in order to compare to minimum detection levels of existing and future instruments, the simulation results can be normalized energetically and then calibrated for a given specific instrument (this is going to be presented elsewhere). 

The integrated (across the spectrum used) energy emitted, per unit time, through neutrinos, is presumed to be a fraction of the fast (non-thermal) proton power in the jet during the eruption modelled. The latter energy, is, in turn, a fraction of the total jet kinetic power, or kinetic luminocity, $L_k$. For $L_k = 10^{40}$ erg/s, then $L_{fp} = 10^{36}$ erg/s, if $f_{fp} = 10^{-4}$. Assuming now a neutrino fraction of $fn = 0.5$ we then obtain $L_n = 0.5 \times 10^{36}$ erg/s.

In general, to scale the neutrino intensity, a normalization factor F$_{norm}$ is needed. In our case, that factor results from energetic arguments. This factor should multiply the neutrino emissivity of Fig. 4 to obtain the neutrino intensity along the jet. As an example, assuming, as above, a fast proton energy fraction of 10$^{-4}$, by equating the area under the $\rho_{1}$ curve of Fig. 4 ($\rho_{1}$ corresponds to the average bulk proton density at the jet base) to the fast proton fraction of the jet kinetic luminosity L$_{k}=10^{40}$ ergs/s, we obtain F$_{norm}$ approximately equal to F$_{norm}$=10$^{54}$ erg/GeV. The latter corresponds to a maximum neutrino intensity of about 0.5$\times$ 10$^{32}$, which is compatible with the results of Ref. \cite{Reynoso09} (see e.g fig. 8 of this reference). We note that, even though our model is quite detailed dynamically, its level of detail cannot be fully compared to observations as of today. The reason is that current and upcoming terrestrial neutrino detectors cannot resolve that much detail, due to the distance of the microquasars from Earth. Therefore, when compared to observations, the predictive power of our model is not very different from that of simpler models \cite{Reynoso_2008,Reynoso09}.

\begin{table}[ht]
\begin{tabular}{|c|cccccccccc|}
\hline\hline
  &  &  &  &  &  &  &  &  &  & \\ 
 &\multicolumn{10}{|c|}{Number density (in $10^{10}$ proton/cm$^3$)} \\ 
 \hline
   &  &  &  &  &  &  &  &  &  & \\
& $\rho_1$ & $\rho_2$ & $\rho_3$ & $\rho_4$ & $\rho_5$ & 
 $\rho_6$ & $\rho_7$ & $\rho_8$ & $\rho_9$ & $\rho_{10}$ \\ 
  \hline
   &  &  &  &  &  &  &  &  &  & \\ 
 $\rho_j = $ & 124.221 & 33.152 & 29.530 & 28.786 & 
      29.220 & 36.558 & 11.627 &  0.260 & 0.142 & 0.138 \\ 
\hline\hline
\end{tabular}
\caption{Number density values $\rho_{j}, j=1,2,\ldots$ averaged over ten slices cut as grid cross sections perpendicular to the jet axis, along the jet axis. The densities are sampled at intervals of 2.0$\times 10^{11}$ cm, to a total grid length along the jet axis direction of 2.0$\times 10^{12}$ cm.}
\label{table_density}
\end{table}

We should point out that, in order to convert quantities from the jet reference frame to our rest frame, the calculational procedure can be e.g.~that of \cite{PS_2001} or that of \cite{TR_2011}. In the present work, we apply the treatment of \cite{PS_2001}. The jet direction has been incorporated as a global effect within the jet, by imposing a fixed angle between the velocity direction of the flow and the line of sight to an observer here on Earth. Furthermore, the jet flow speed is taken to be set to 0.26c, the average flow estimated for the jet. In addition, the line-of-sight direction is assumed constant all over the jet, at an angle of $\theta$=78 degrees, to the jet axis. This was done to keep the calculations within limits. In principle, each computational cell may have a different setting for the angle between its local velocity and the line of sight, as well as for the local emission calculation performed using its localized velocity value. In both cases, a much longer computational time is required.

Before closing, it is worth mentioning that, the total power emitted from the jet obtained in the present work is only a first approximation to the intensity estimate. Consequently, a more detailed comparison to detectors is required. Our model is indeed able to provide, for a given direction to the observer, individual Doppler effects for each 3D computational cell, and then integrate them numerically (see Ref. \cite{Smponias_2017}).


\section{Summary and Conclusions}

In the present work, we evaluated the emissivity of neutrinos originating from hadronic MQ jets, 
where p-p collisions occur at shock fronts, leading to cascades of secondary particles, culminating 
to neutrino emission. We have implemented a new model describing the mass distribution along the jet 
axis, using the PLUTO relativistic magneto-hydrodynamic (RMHD) code (hydrocode). More specifically, 
the PLUTO code was executed incorporating a toroidal magnetic field component in the jet, resulting 
to a confined jet structure, the degree of confinement depending on the value of the field.
For each cross section slice, cut along the jet (perpendicular to the jet axis) we calculated the 
mean values of the mass density. Then, we proceeded, in this manner, to process a number of $100$ 
slices, covering the spatial range from the jet base to the end of the computational grid. 

The main conclusion extracted from this analysis was that the hydrocode model (not based on explicit geometrical assumptions), employed for the hadronic jet, is dynamically a realistic tool. This is why we decided to utilise the PLUTO code for dynamical calculations as a basis for further investigation of the neutrino and gamma-ray emissivities from the jet. For our present calculations, we used the semi-analytic
approach, in order to estimate the neutrino emissivity, as described in our previous
work.

In studying the neutrino emissivity per grid cell, we set up a model geometry reminiscent of 
the semi-analytical method, but using the PLUTO hydrocode, while employing the known radiative 
formalism as discussed in the Introduction. This computational 
tool has previously provided us with a realistic
modelling of radio and gamma-ray emission and in this work with efficient estimation of neutrino emission events originating from micro-quasar jets. For the observation of such neutrino fluxes, current terrestrial detectors (e.g. IceCube at South Pole) are in operation.

\section{Conflict of Interests}

The authors declare that there is no conflict of interests regarding the publication of this paper.

\end{document}